\documentclass[prd,superscriptaddress]{revtex4}
\usepackage{amsmath}
\usepackage{graphicx}
\usepackage{bm}

\usepackage{amssymb}

\def\gsim{\;\rlap{\lower 2.5pt
 \hbox{$\sim$}}\raise 1.5pt\hbox{$>$}\;}
\def\lsim{\;\rlap{\lower 2.5pt

   \hbox{$\sim$}}\raise 1.5pt\hbox{$<$}\;}

\begin{document}

\def\be{\begin{equation}}
\def\ee{\end{equation}}
\def\ba{\begin{eqnarray}}
\def\ea{\end{eqnarray}}
\def\nn{\nonumber}
\def\trh{T_{\rm{RH}}}
\def\k0{k_0^{\rm{p}}}

\title{Using pulsar timing arrays and the quantum normalization condition to constrain relic gravitational waves}

\author{M.L. Tong}
\thanks{Email: mltong@ntsc.ac.cn}
\affiliation{National Time Service Center, Chinese Academy of Sciences, Xi'an, Shaanxi 710600,  China }
\affiliation{Key Laboratory of Time and Frequency Primary Standards, Chinese Academy of Sciences, Xi'an, Shaanxi 710600,  China}
\author{Y. Zhang}
\thanks{Email: yzh@ustc.edu.cn}
\affiliation{Key Laboratory of Galactic and Cosmological Research,
        Center for Astrophysics, University of Science and Technology of China, Hefei, Anhui, 230026,  China}
\author{W. Zhao}
\affiliation{Key Laboratory of Galactic and Cosmological Research,
        Center for Astrophysics, University of Science and Technology of China, Hefei, Anhui, 230026,  China}

\author{J.Z. Liu}
\affiliation{Xinjiang Observatory,  National Astronomical Observatories£¬Chinese Academy of Sciences£¬150, Science 1-Street Urumqi, Xinjiang 830011 China
}
\author{C.S. Zhao}
\affiliation{National Time Service Center, Chinese Academy of Sciences, Xi'an, Shaanxi 710600,  China }
\affiliation{Key Laboratory of Time and Frequency Primary Standards, Chinese Academy of Sciences, Xi'an, Shaanxi 710600,  China}

\author{T.G. Yang}
\affiliation{National Time Service Center, Chinese Academy of Sciences, Xi'an, Shaanxi 710600,  China }
\affiliation{Key Laboratory of Time and Frequency Primary Standards, Chinese Academy of Sciences, Xi'an, Shaanxi 710600,  China}

\begin{abstract}

In the non-standard model of relic gravitational waves (RGWs) generated in the early universe,
the theoretical spectrum of
is mainly described by an amplitude $r$ and a spectral index $\beta$,
the latter usually being determined by the slope of the inflation potential.
Pulsar timing arrays (PTAs) data have imposed constraints
on the amplitude of strain spectrum
for a power-law form as a phenomenological model.
Applying these constraints to
a generic, theoretical spectrum with $r$ and $\beta$
as independent parameters,
we convert the PTAs constraint into an upper bound on the index $\beta$,
which turns out to be
less stringent than those upper bounds from BBN, CMB, and LIGO/VIRGO,
respectively. Moreover, it is found that PTAs constrain the non-standard RGWs
more stringent than the standard RGWs. 
If the condition of the quantum normalization is imposed
upon a theoretical spectrum of RGWs,
 $r$ and $\beta$ become related.
With this condition,
a minimum requirement of the horizon size during inflation
is greater than the Planck length
results in an upper bound on $\beta$,
which is comparable in magnitude to that by PTAs.
When both PTAs and the quantum normalization are applied to
a theoretical spectrum of RGWs,
constraints can be obtained for other cosmic processes of the early universe,
such as the reheating, a process less understood
observationally so far.
The resulting constraint is consistent with
the slow-roll, massive scalar inflation model.
The future  SKA
will be able to constrain RGWs further
and might even detect RGWs,
rendering an important probe to the very early universe.

\

\end{abstract}

\maketitle

\large

\section{ Introduction}

A stochastic background of relic
gravitational waves (RGWs) is a natural prediction
of general relativity and quantum mechanics
\cite{grishchuk1,grishchuk,grishchuk3,grishchuk05,starobinsky,Maggiore,Giovannini}.
As fluctuations of the metric of spacetime,
RWGs could be originated from the quantum fluctuations during the inflationary stage.
Since their interaction with other cosmic components was typically very weak,
RGWs, after being generated,
are determined by the expanding behavior of spacetime background
and only slightly modified
by other cosmic processes during the evolution \cite{zhang2,Zhang4,Miao,Schwarz,SWang}.
Thus RGWs carry a unique information of the early universe
and serve as a probe into the universe much
earlier than the cosmic microwave background (CMB).
As an important feature for detection purpose,
RGWs exist everywhere and all the time,
and the spectrum spreads a very broad range of frequency, $10^{-18}- 10^{10}$ Hz,
constituting one of the major scientific targets of various types of GW detectors,
including the ground-based interferometers,
such as  LIGO \cite{ ligo1},
VIRGO \cite{virgo}, GEO \cite{geo} and KAGRA \cite{KAGRA}
at the frequency range $10^2-10^3$ Hz;
the space interferometers, such as the future eLISA/NGO \cite{lisa},
DECIGO \cite{decigo,Kawamura}, and BBO \cite{Crowder,Cutler}
at the frequencies $10^{-4}-10^0$ Hz;
the waveguide detector \cite{cruise},
the proposed gaussian maser beam detector around GHz \cite{fangyu},
and the 100 MHz detector with a pair of interferometers \cite{Akutsu}.
Furthermore,  the very low frequency portion of
RGWs also contribute to the CMB anisotropies
and polarizations  \cite{basko},
yielding a distinguished magnetic type of polarization of CMB,
 which has been a detecting goal of CMB observations,
such as WMAP \cite{page,Komatsu2009,Komatsu,Hinshaw12},
Planck \cite{Planck,Planck2},  and the proposed CMBpol \cite{CMBpol}.

Another important tool to detect RGWs
is the  pulsar timing arrays (PTAs) \cite{Sazhin,Kaspi}.
The detection of the lower frequency limit is the inverse of the  observation time span, $\sim 10^{-9}$ Hz, and the upper frequency limit corresponds to the observation time interval, $\sim 10^{-7}$ Hz.
By correlating the pulse arrival timings
of an array of selected millisecond pulsars,
one can, in principle, disentangle the signal of gravitational waves from
the timing data of a long period of observations.
Currently, there are several such detectors running,
such as the Parkes Pulsar Timing Array (PPTA) \cite{PPTA},
European Pulsar Timing Array (EPTA) \cite{Haasteren},
the North American Nanohertz Observatory for Gravitational Waves (NANOGrav) \cite{Demorest},
and the  much more sensitive Five-hundred-meter Aperture Spherical Radio Telescope (FAST) \cite{Nan} and  Square Kilometre Array (SKA) \cite{Kramer}
are also under planning.
The typical response frequency of PTA is nanoHertz,
inversely proportional to the observation period.
In this rage of frequencies,
both RGWs and the gravitational radiation by supermassive black hole binaries
\cite{Jaffe,Sesana,Yardley,Liujz,WuZhang}
are the major of scientific targets of PTAs.
Besides, PTAs can be instrumental in study of cosmology \cite{WZhao}.

Although RGWs has not been detected directly so far,
various constraints on RGWs have been studied.
The successful Big Bang nucleosynthesis (BBN)
puts a tight upper bound on
the total energy fraction $\Omega_{gw}h^2 < 7.8\times 10^{-6}$
of gravitational waves for frequencies $>10^{-10}$Hz \cite{Allen1999,Allen1996}.
Besides,
CMB + galaxy surveys + Lyman-$\alpha$
also  yields a similar bound on $\Omega_{gw}h^2 < 6.9\times 10^{-6}$
for extended lower frequencies $>10^{-15}$Hz \cite{Smith2006}.
The RGWs  spectrum is,  to a large extent,
prescribed by the initial amplitude $r$ (the tensor-scalar ratio),
the spectral index  $\beta$,  and the running index $\alpha_t$.
For scalar field inflationary models,
while $r$ is largely determined by the energy scale of inflaton potential,
$\beta$ is determined by the slope,
and $\alpha_t$ by the curvature of the potential \cite{KosowskyTurner,Smith2}.
So the indices $\beta$ and $\alpha_t$ are more powerful in discriminating
inflationary models.
By integrating the spectrum of RGWs,
the afore-mentioned bounds  have been converted into
the constraints on $\beta$ and  $\alpha_t$ for fixed $r$
\cite{TongZhang,zhangtong}.
The WMAP observations of the spectra of CMB anisotropies and polarization
have yielded upper bounds on
the  ratio  $r$ of RGWS
for the fixed scalar index \cite{Komatsu2009,Komatsu}.
The observational data of LIGO/VIRGO has led to constraint
on  $\beta$ and  $\alpha_t$ of RGWs
at fixed $r$  \cite{TongZhang}.
Based on the LIGO S5 data,
the signal-noise ratios for $\beta$ and  $\alpha_t$
haven been obtained by correlating
the given pair of detectors \cite{zhangtong}.

The amplitude of RGWs at frequencies $\sim 10^{-9}$Hz of PTAs
is about ten orders higher than
at frequencies $10^2-10^3$ Hz of LIGO, VIRGO, etc \cite{zhang2,Zhang4}.
One might expect to get tighter constraints on RGWs thereby.
Recently, we discussed the constraints and detection  of the RGWs in the
standard hot big bang cosmological model by various PTAs and the future FAST and SKA \cite{zhaozhang}.
In this paper, employing the data of the current and future PTAs,
we will  give the constraints on the non-standard RGWs model \cite{grishchuk1,grishchuk,grishchuk3,grishchuk05} which contains a reheating (or preheating \cite{Tong6}) process occurred  after the end of inflation and before the beginning of the radiation dominated stage of the universe. From the constraints, we also try to study the expansion behavior and the physical processes happened in the very early universe.
To be as general as possible,
we take $r$ and $\beta$ as two free parameters of RGWs,
and do not include $\alpha_t$ for simplicity.
Specifically,
we will focus on the constraint on the spectral index $\beta$,
which is also the power-law index of cosmic expansion during the inflation,
determined by the specific inflation models.
PTAs have been used to constrain the GW background generated by the cosmic strings
 \cite{Sanidas},
which  has a different origin and different spectral features from RGWs.

While $\beta$ is predicted by the potential
in specific inflation models,
$r$ could be determined by certain extra condition,
such as the consistency condition \cite{LiddleLyth}.
In  regard to this issue,
there is another kind of condition,
the so-called quantum normalization
of amplitude of RGWs \cite{grishchuk}.
One can treat the RGWs field  $h_{ij}$ as a quantum field in the vacuum state
 when initially generated,
require that each mode $k=\omega/2\pi$ have
an energy $\frac{1}{2}\hbar \omega$.
This leads to the quantum normalization for
the initial condition of RGWs,
in which $r$ and $\beta$ are no longer independent.
In this paper, we will also use the theoretical condition of  quantum normalization to constrain $\beta$ for given values of $r$, complementary to the observational constraints from PTAs.

Finally,
when both PTAs and the quantum normalization are applied on RGWs,
instead of the inflation expansion via $r$ and $\beta$,
a constraint can be obtained upon the reheating process,
which  is the least understood,
 theoretically as well as observationally,
among the cosmic processes  so far.

We neglect the effect on the spectrum of RGWs
caused by the neutrino free-streaming \cite{Miao,Weinberg,yuki},
since its modifications just fall out of the band of frequencies of PTAs.
As for the QCD transition and the $e^+ e^-$ annihilation \cite{Schwarz,SWang},
their modifications on the spectrum of RGWs
occur for $f>10^{-9}$ Hz and $f>10^{-12}$ Hz, respectively,
within the band of frequencies of PTAs.
But the combined result is only a small reduction of amplitude of RGWs
by $\sim 30\%$,
which can be simply absorbed into the definition of the amplitude $r$
in our treatment.
In this paper we use unit with $c=\hbar=k_B=1$.

\section{   RGWs in the accelerating universe}

In a spatially flat universe, the general  Friedmann-Robertson-Walker metric  is
\be
ds^2=a^2(\tau)[-d\tau^2+(\delta_{ij}+h_{ij})dx^idx^j],
\ee
where $a(\tau)$ is the scale factor, $\tau$ is the conformal time, and $h_{ij}$ stands for the  perturbations to the  homogenous and isotropic spacetime background. In general, there are three kinds of perturbations: scalar perturbation, vectorial perturbation and tensorial perturbation.
Here we are only interest in  the tensorial perturbation, i.e., gravitational waves.  In the transverse-traceless (TT) gauge,
$h_{ij}$ satisfies:   $\frac{\partial h_{ij}}{\partial x^j}=0$  and $h^i_{\,\,i}=0$, where we used the Einstein  summation convention. With the evolutions of the cosmic background, RGWs satisfy
\be\label{weq}
\partial_\mu(\sqrt{-g}\partial^\mu h_{ij}(\tau,{\bf{x}}))=0,
\ee
according to Einstein field equation, where $g\equiv {\rm det}(g_{\mu\nu})$.
In the Fourier $k$-modes space, the general solution of Eq. (\ref{weq}) is given by
\be
\label{planwave}
h_{ij}(\tau,{\bf x})=
   \sum_{\sigma}\int\frac{d^3\bf{k}}{(2\pi)^{3/2}}
         \epsilon^{(\sigma)}_{ij}h_k^{(\sigma)}(\tau)e^{i\bf{k}\cdot{x}},
\ee
where $\sigma=+,\times$ stands for the two polarization states,
 the comoving wave number $k$ is related with
the wave vector $\mathbf{k}$ by $k=(\delta_{ij}k^ik^j)^{1/2}$,
$h_{-k}^{(\sigma)*}(\tau)=h_k^{(\sigma)}(\tau)$
ensuring  $h_{ij}$ be real, and the polarization
tensor  $\epsilon^{(\sigma)}_{ij}$ satisfies \cite{grishchuk}:
\be
\epsilon^{(\sigma)}_{ij}\epsilon^{(\sigma'){ij}}=2\delta_{\sigma\sigma'}, \quad
\epsilon^{(\sigma)}_{ij}\delta^{ij}=0, \quad
\epsilon^{(\sigma)}_{ij}n^j=0,\quad
\epsilon^{(\sigma)}_{ij}(-\mathbf{k})=\epsilon^{(\sigma)}_{ij}(\mathbf{k}).
\ee
In terms of the mode $h^{(\sigma)}_{k}$,
the wave equation is
\be \label{eq}
h^{ (\sigma) }_{k}{''}(\tau)
+2\frac{a'(\tau)}{a(\tau)}h^{ (\sigma) }_k {'}(\tau)
+k^2 h^{(\sigma)}_k(\tau )=0,
\ee
where a prime means taking derivative with respect to  $\tau$.
The two polarizations of
$h^{(\sigma)}_k(\tau )$ have the same statistical
properties and give equal contributions to the unpolarized RGWs background,
so the super index $(\sigma)$ can be dropped.
For a power-law form of $a(\tau) \propto \tau^\alpha$,
Eq. (\ref{eq}) has an analytic
solution which is a linear combination of
Bessel and Neumann functions:
\be \label{hom}
h_k(\tau)=\tau^{\frac{1}{2}-\alpha}
 \big[C_1 J_{\alpha-\frac{1}{2}}(k \tau)
      +C_2   N_{\alpha-\frac{1}{2}}(k \tau)\big],
\ee
where the constants $C_1$ and $C_2$ for each stage are determined
by the continuities of $h_k(\tau)$ and  $h'_k(\tau)$
at the joining points
$\tau_1,\tau_s,\tau_2$ and $\tau_E$ \cite{zhang2,Zhang4,Miao}
for the different stages of the universe.
The scale factor in a series of cosmic expansion stages can be written
in power-law forms \cite{grishchuk,Miao,TongZhang,Tong6}
as the following:

The inflationary stage:
\be \label{inflation}
a(\tau)=l_0|\tau|^{1+\beta},\,\,\,\,-\infty<\tau\leq \tau_1,
\ee
where the inflation index $\beta$
is a model parameter describing the expansion behavior of inflation.
The special case of $\beta=-2$  corresponds the exact de Sitter expansion
driven by a constant vacuum energy density.
However, for inflationary expansions driven by some dynamic field,
the predicted values of $\beta$ scatter around  $-2$,
depending on specific models.
In the single-field
slow-roll inflation model, one always has $\beta<-2$, i.e., red spectrum \cite{LiddleLyth,Tong6,Tong8}.
 However, some other inflation
models, such as the phantom inflations \cite{Piao} also
predict the blue spectrum, which has not been excluded
by observations \cite{Stewart,Camerini}.
Besides,  a relation   $n_s=2\beta+5$ with  $n_s$ being the
the scalar spectral index
of primordial perturbations, has been often employed \cite{grishchuk3,TongZhang}.
The observed result of CMB isotropies by WMAP \cite{Komatsu2009,Komatsu} indicates
the scalar spectral index $n_s \simeq 0.96$,
corresponding to $\beta\simeq -2.02$.
In this paper we mainly focus on $\beta$ as a major free parameter of RGWs in analysis.

The reheating stage :
\be\label{betas}
a(\tau)=a_z|\tau-\tau_p|^{1+\beta_s},\,\,\,\,\tau_1\leq \tau\leq \tau_s,
\ee
where the parameter  $\beta_s$ describes
the expansion behavior of the preheating stage
from the end of inflation to the happening of
reheating process followed by the radiation-dominant stage.
As shown in \cite{Tong8},
$\beta_s$ only affects the RGWs in very high frequencies.
In this paper, we will take the value $\beta_s=1$ \cite{Tong6,Starobinsky2,Kuroyanagi},
since we focus on the very-low frequency bands $10^{-9}-10^{-7}$ Hz
for PTAs.
Compared with the inflation,
the reheating as a dynamical process is less understood so far,
either theoretically or observationally.
In later part of this paper,
we will also try to give some information of
the increase of scale factor $\zeta_1\equiv{a(\tau_s)}/{a(\tau_1)}$
during the reheating stage.

The radiation-dominant stage :
\be \label{r}
a(\tau)=a_e(\tau-\tau_e),\,\,\,\,\tau_s\leq \tau\leq \tau_2.
\ee

The matter-dominant stage:
\be \label{m}
a(\tau)=a_m(\tau-\tau_m)^2,\,\,\,\,\tau_2 \leq \tau\leq \tau_E.
\ee

The accelerating stage up to the present time $\tau_0$:
\be \label{accel}
a(\tau)=l_H|\tau-\tau_a|^{-\gamma},\,\,\,\,\tau_E \leq \tau\leq
\tau_0,
 \ee
where $\gamma\simeq 2$ for the energy density contrast
$\Omega_{\Lambda}\simeq0.7 $  \cite{zhangtong}.
Conveniently,
one  chose the normalization  $|\tau_0-\tau_a|=1$ \cite{zhang2,Zhang4}, i.e.,
the present scale factor $a(\tau_0)=l_H$.
By  definition,
one has $l_H=\gamma/H_0$,
where the Hubble constant $H_0=100\, h$ km s$^{-1}$ Mpc$^{-1}$
with  $h= 0.673$.

All the constants referring from Eq.(\ref{inflation}) to Eq.(\ref{accel})
can be determined by the continuity of $a(\tau)$ and $a'(\tau)$
at the four given joining points $\tau_1$, $\tau_s$, $\tau_2$ and $\tau_E$,
which are equivalent to the four given increases of
the scale factor:
$\zeta_1= {a(\tau_s)}/{a(\tau_1)}$,
$\zeta_s\equiv{a(\tau_2)}/{a(\tau_s)}$,
$\zeta_2\equiv{a(\tau_E)}/{a(\tau_2)}$,
and
$\zeta_E\equiv{a(\tau_0)}/{a(\tau_E)}$.
For the accelerating stage in the simple $\Lambda$CDM model,
one  has
$\zeta_E =1+z_E \simeq ({\Omega_\Lambda}/{\Omega_m})^{1/3}\simeq1.3$,
where $z_E$ is the redshift when the  accelerating expansion begins.
For the matter-dominated stage,
one has  $\zeta_2 =\frac{a(\tau_0)}{a(\tau_2)} \frac{a(\tau_E)}{a(\tau_0)}
=(1+z_{eq}) \zeta_E^{-1}$ with $z_{eq}=3402$ \cite{Planck2}.

For the radiation stage,
the value of $\zeta_s$ depends on
the reheating temperature $\trh$,
at which the radiation stage begins.
In the Big Bang cosmology,
following the inflationary expansion
is the reheating process that converts the vacuum energy into radiation.
This process is not yet well understood,
either observationally or theoretically.
Associated with this issue
is the uncertainty of $\trh$.
Due to the conservation of the entropy, the increase of
the scale factor during the radiation-dominated era
can be written in terms of $\trh$ \cite{Tong6,Tong8}:
\be\label{delta1}
\zeta_s=\frac{T_{\rm{RH}}}{T_{\rm{CMB}}(1+z_{eq})}
          \left(\frac{g_{\ast s}}{g_{\star s}}\right)^{1/3},
\ee
where   $T_{\rm{CMB}}=2.725\, {\rm K} =2.348\times10^{-13}$ GeV is the
present CMB temperature,
$g_{\ast s}\simeq200$ is the effective number of relativistic species contributing
to the entropy after the reheating,
and  $g_{\star s}=3.91$ is the one after recombination  \cite{yuki,Tong6}.
For the single field inflation, CMB data would yield
 the lower bound of $\trh \gtrsim 6\times10^3$ GeV,
and the most upper bound  could be up
to  $\trh \lesssim3\times10^{15}$ GeV  \cite{martin}.
The slow-roll massive scalar field inflation
would predict $\trh = 5.8\times 10^{14}$ GeV \cite{Tong6}.
In the supersymmetry scenarios,
gravitinos production would give an upper bound
$\trh  \lesssim10^8$ GeV \cite{gravitinos}.
Thus, for our purpose,
we will consider the range $\trh \sim(10^4-10^8)$ GeV.

For the reheating process,
the parameter  $\zeta_1$ is also uncertain.
Based on the slow-roll scalar inflation models \cite{Mielczarek,Tong6,Tong8},
 $\zeta_1$ depends on the specific form of the potential $V$ that drives the inflation.
When calculating the spectrum of RGWs in low frequencies, we just choose some particular values of $\zeta_1$, as it only affects RGWs in very high frequencies which will be shown below.  Apart from that,  we  will treat  $\zeta_1$ as a parameter of the reheating,
and put certain constraints on it
by a combination of PTAs data
and the condition of the quantum normalization of RGWs.

The spectrum of RGWs $h(k,\tau)$ is defined by
\be \langle
h^{ij}(\tau,\mathbf{x})h_{ij}(\tau,\mathbf{x})\rangle\equiv\int_0^\infty
h^2(k,\tau)\frac{dk}{k},
\ee
where the angle brackets mean ensemble average.
The dimensionless spectrum $h(k,\tau)$ relates to the mode $h_k(\tau)$ as
\cite{TongZhang}
\be \label{relation0}
h(k,\tau)=\frac{\sqrt{2}}{\pi}k^{3/2} |h_k(\tau)|.
\ee
At the present time $\tau_0$ the above gives
the present RGWs spectrum $h(k,\tau_0)$.
Assuming that the wave mode crosses the  horizon of the universe when
$\lambda/(2\pi)=1/H$, then the characteristic comoving wave number
at a certain joining time $\tau_x$ can be defined as
\be\label{wavenumber}
k_x\equiv k(\tau_x) =  a(\tau_x) H(\tau_x),
\ee
which is little different from Ref.\cite{Tong6}. For example, the characteristic comoving wave number at present is
 $k_H=a(\tau_0)H_0=\gamma$. By a similar calculation,
one has and the following relations:
\be\label{frelation}
  \frac{k_E}{k_H}
    = \zeta_E^{-\frac{1}{\gamma}},\quad
    \frac{k_2}{k_E}=\zeta_2^{\frac{1}{2}},\quad
    \frac{k_s}{k_2}=\zeta_s, \quad
    \frac{k_1}{k_s}=\zeta_1^{\frac{1}{1+\beta_s}}.
 \ee
In the present universe,
the physical frequency relates to a comoving wave number  $k$ as
\be \label{freq}
f=  \frac{k}{2\pi a (\tau_0)} = \frac{k}{2\pi l_H}.
\ee

 The present energy density contrast of RGWs defined by
$ \Omega_{GW}=\langle\rho_{g}\rangle/{\rho_c}$,
where $\rho_g=\frac{1}{32\pi G}h_{ij,0}h^{ij}_{,0}$
is the energy density of RGWs
and $\rho_c=3H_0^2/8\pi G$ is the critical energy density,
and is given by
 \cite{grishchuk3,Maggiore}
\be\label{gwe}
\Omega_{gw}=
\int_{f_{low}}^{f_{upper}} \Omega_{g}(f)\frac{df}{f},
\ee
with
\be\label{omega}
\Omega_{g}(f)=\frac{2\pi^2}{3}
        h^2_c(f)
     \Big(\frac{f}{H_0}\Big)^2
\ee
being the dimensionless  energy density spectrum.
Here we have used  a  notation, $h_c(f)\equiv h(f,\tau_0)/\sqrt{2}$,
called  the {\it characteristic strain spectrum} \cite{Maggiore}
or {\it chirp amplitude} \cite{Boyle}.
The lower and upper limit of integration in Eq.(\ref{gwe})
can be taken to be  $f_{low}\simeq f_E$
and  $f_{upper}\simeq f_1$, respectively,
since only the wavelength of the modes
inside the horizon contribute to the total energy density.

The analytic solutions were studied by many authors
\cite{Zhang4,yuki,Miao,Kuroyanagi,TongZhang}.
For simple discussions but without losing generality,
in this paper, we employ the approximate solutions of RGWs listed in \cite{Tong6}:
\ba\label{htaue}
&&h(k,\tau_0)=A\left(\frac{k}{k_H}\right)^{2+\beta}, \qquad k\leq k_E;\\
&&h(k,\tau_0)=A\left(\frac{k}{k_H}\right)^{\beta-\gamma}(1+z_E)^{-\frac{2+\gamma}{\gamma}}, \qquad  k_E\leq k\leq k_H;\\ \label{htauh}
&&h(k,\tau_0)=A\left(\frac{k}{k_H}\right)^{\beta}(1+z_E)^{-\frac{2+\gamma}{\gamma}}, \qquad  k_H\leq k\leq k_2;\\\label{hpulsar}
&&h(k,\tau_0)=A\left(\frac{k}{k_H}\right)^{1+\beta}\left(\frac{k_H}{k_2}\right)(1+z_E)^{-\frac{2+\gamma}{\gamma}}, \qquad  k_2\leq k\leq k_s;\\ \label{hf1}
&&h(k,\tau_0)=A\left(\frac{k}{k_H}\right)^{1+\beta-\beta_s}\left(\frac{k_s}{k_H}\right)^{\beta_s}\left(\frac{k_H}{k_2}\right)
(1+z_E)^{-\frac{2+\gamma}{\gamma}}, \,\,  k_s\leq k\leq k_1.
\ea
where the coefficient $A$ can be determined by the initial condition.
After all it should be determined by observations.
We will discuss this issue below.

The amplitude of RGWs
at a pivot wave number $k^p_0=k_0/a(\tau_0)=0.002$ Mpc$^{-1}$ \cite{Komatsu}
can be normalized to the tensor-to-scalar ratio
\cite{Peiris,Spergel07}:
\be\label{ratio}
r\equiv\frac{\Delta^2_h(k_0)}{\Delta^2_{\mathcal{R}}(k_0)},
\ee
where $\Delta^2_h(k_0)\equiv h^2(k_0,\tau_0)$ \cite{Tong6}
and $\Delta^2_{\mathcal{R}}(k_0)=2.427\times10^{-9}$ given
by WMAP\,9+BAO+$H_0$ \cite{Hinshaw12}.
At present only
observational constraints on $r$ have been given. The upper bounds of $r$ are
 constrained by WMAP\,9+eCMB+BAO+$H_0$\cite{Hinshaw12}  as  $r<0.13$ for the vanishing scalar running spectral index $\alpha_s$ and $r<0.47$ for the non-vanishing $\alpha_s$, respectively.
  More tighter constraints of $r$ were given by Planck+WMAP \cite{Planck2}: $r<0.11$ and $r<0.26$ for
  the  vanishing $\alpha_s$ and the  non-vanishing $\alpha_s$, respectively. In this paper, we will follow the constraints given in \cite{Planck2}.  On the other hand, using a discrete,
model-independent measure of the degree
of fine-tuning required, if $0.95\lesssim n_s<0.98$,
in accord with current measurements,
the tensor-to-ratio satisfies $r\gtrsim10^{-2}$  \cite{Boyle}.
We will take $r$ lying in the range of $(0.01,0.26)$
in our demonstrations.
Since  $k_H\leq k_0\leq k_2$, from Eq. (\ref{htauh}) one has
\be\label{primordial}
h(k_0,\tau_0)=A\left(\frac{k_0}{k_H}\right)^{\beta}
            (1+z_E)^{-\frac{2+\gamma}{\gamma}}
=[\Delta^2_{\mathcal{R}}(k_0)r]^{1/2},
\ee
telling that  $A$ can be determined
for the given ratio  $r$ and the index $\beta$.

\begin{figure}
\resizebox{100mm}{!}{\includegraphics{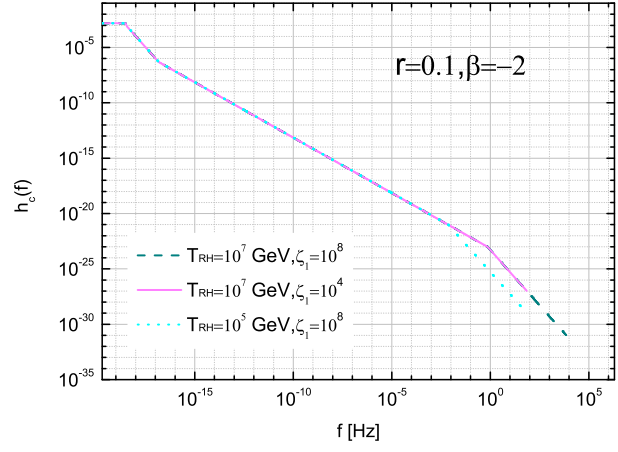}}
\caption{\label{hcapprox}
The characteristic strain spectra $h_c(f)$
of RGWs for various values of $\trh$ and $\zeta_1$
at fixe $r=0.1$ and $\beta=-2$ for demonstration.
}
\end{figure}

In Fig.\ref{hcapprox}, we plot the characteristic
strain spectrum $h_c(f)$ of RGWs for various  values of
$\trh$ and $\zeta_1$ for the fixed $r=0.1$ and $\beta=-2$.
One can see that, the variations of $\trh$ and $\zeta_1$
affect the RGWs only at frequencies $f>10^{-1}$Hz,
far away from the frequency window $\sim10^{-9}-10^{-7}$ Hz of PTAs.
So, for demonstration below, we choose $\trh=10^7$ GeV and $\zeta_1= 10^{8}$
allowed by the slow-roll inflation models \cite{Mielczarek,Tong6,Tong8}.

\begin{figure}
\resizebox{100mm}{!}{\includegraphics{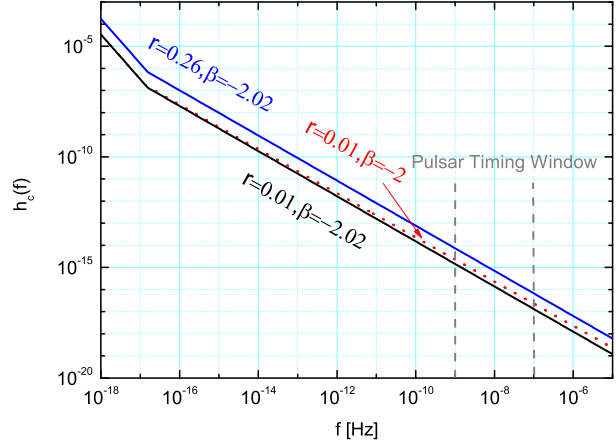}}
\caption{\label{hcapprox2}
The characteristic strain spectra of
RGWs for various values of $r$ and $\beta$
without considering the quantum normalization.
The vertical dashed lines stand for the detecting window of PTAs.
}
\end{figure}

Unlike $\trh$ and $\zeta_1$,
small variations of the parameters $r$ and $\beta$ do
significantly affect the spectrum in all frequencies.
In Fig.\ref{hcapprox2},
we show $h_c(f)$ for $\beta=-2$ of the exact de Sitter expansion,
and for $\beta=-2.02$ corresponding to the scalar spectral index $n_s=0.96$
as given by Planck \cite{Planck2}.
A greater $r$ leads to greater amplitudes for
the whole frequency range,
while a greater $\beta$
 leads to greater amplitudes at higher frequencies.

\section{ Constraints by pulsar timing arrays}

The existence of gravitational waves
 will change the geodesic of the photons  from mili-second
 pulsars to the observer.
Consequently,
the times of arrival (TOAs) of the electromagnetic signals from pulsars
 will be perturbed, forming the so-called timing residuals \cite{Sazhin}.
If the gravitational waves are strong enough,
one could extract their signals buried in
the data of the timing residual measurements.
Even RGWs are very weak,
still,
constraints on the amplitude of GWs can be  obtained
from the long-time accumulating data of timing residuals.

Over the last 30 years various
data from PTA experiments have set constraints
\cite{Bertotti,Kaspi,Thorsett,McHugh,Lommen,Jenet,Hobbs2009,Haasteren,Demorest}.
In the practice of data analysis of PTAs,
the gravitational wave is usually modeled simply with a power-law form
of the characteristic strain spectrum:
 \be\label{hc1}
 h_c(f)=A_1\left(\frac{f}{{\rm yr}^{-1}}\right)^\alpha,
 \ee
where $A_1$ is the amplitude and $f$ is the frequency in unit ${\rm yr}^{-1}$.
For PTAs the detection frequency band is  $10^{-9}\leq f\leq10^{-7}$ Hz.
Relevant to this band of frequency,
the RGWs mode by our calculation
is given in Eq.(\ref{hpulsar}),
and the corresponding, theoretical characteristic strain spectrum
has the following form:
\be\label{chara}
h_c(f)=\frac{A}{\sqrt{2}}
\left(\frac{f}{f_H}\right)^{1+\beta}
\left(\frac{f_H}{f_2}\right)(1+z_E)^{-\frac{2+\gamma}{\gamma}},
\ee
where $f_H=H_0/(2\pi)=3.47\times10^{-19}$ Hz and $f_2=1.56\times10^{-17}$ Hz
due to Eq.(\ref{frelation}).
With the help of Eq. (\ref{primordial}),
Eq. (\ref{chara}) can be rewritten as
\be\label{strain}
h_c(f)=\frac{[\Delta_\mathcal{R}^2(k_0)r]^{1/2}}{\sqrt{2}}
\left(\frac{f_0}{f_2}\right)
\left(\frac{f}{f_0}\right)^{1+\beta},
\ee
where $f_0=3.09\times10^{-18}$ Hz is the pivot frequency.
Note that $f/f_0\gg1$ in the pulsar timing frequency band.
Comparing  Eq.(\ref{hc1}) and Eq.(\ref{strain}) tells that
the power-law index is related to the inflation index via
\be
\alpha=1+\beta.
\ee
In Ref.\cite{zhaozhang}, the power spectrum of RGWs described by the tensor
spectral index $n_t$, which has a relation with $\alpha$,
\be
\alpha=\frac{n_t}{2}-1.
\ee
Then one can straightly  get $n_t=2\beta+4$. $n_t$ was constrained by PTAs
in \cite{zhaozhang}. In this paper, however, we will constrain the parameter $\beta$ instead, which
describes the expansion behavior of the inflation directly.

\begin{figure}
\resizebox{100mm}{!}{\includegraphics{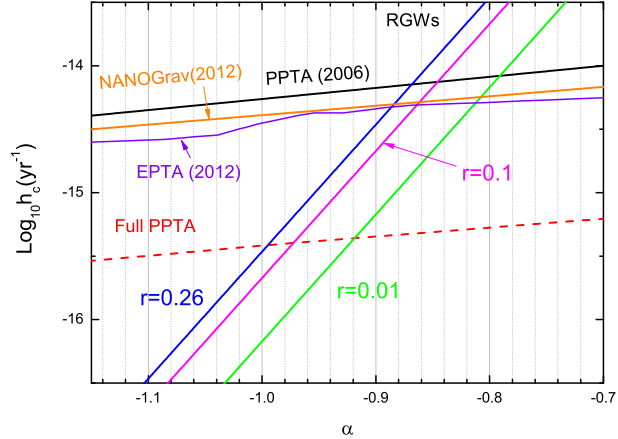}}
\caption{\label{a-alpha}
The constraints of $\alpha $ given by various PTAs
for RGWs with different valus of ratio $r$.
The limit curves of the PPTA (2006)
and the full PPTA are taken from Ref.\cite{Jenet}.
The limit curves of the current EPTA and NANOGrav
are taken from Ref.\cite{Haasteren} and Ref.\cite{Demorest}, respectively.}
\end{figure}

Improving the earlier works \cite{Kaspi,Lommen},
Jenet et al \cite{Jenet} developed a frequentist technique of statistics,
and have placed an upper limit on $A_1$ for each given $\alpha$
in the range $\alpha \in(-2, 1)$.
Expecting a GW stochastic background with a red spectrum
(more power at low frequencies, corresponding to $\alpha <-1$, i.e, $\beta<-2$),
they have selected of
seven pulsars  with formally white spectra
from the observational data of
PPTA and Arecibo experiments.
They also gave the limit curve from the simulated data
of the potential 20 pulsars for the future goal of the PPTA timing
 (see Fig.2 in \cite{Jenet}). From now on, as in Jenet et al \cite{Jenet}, we refer to this simulated data as the ``full PPTA''.
Currently, EPTA \cite{Haasteren}
and  NANOGrav \cite{Demorest}
also gave the similar limit curves.
For investigation,
we quote these upper limit curves of $A_1(\alpha)$ in Fig.\ref{a-alpha}, which are also demonstrated in \cite{zhaozhang}.
For example, at $\alpha=-1$,
the current NANOGrav gives the upper bounds $A_1=4.1\times10^{-15}$,
and
the full PPTA gives $A_1=3.8\times10^{-16}$,  respectively,
at the $95\%$ confidence level.
Note that,
in these results of statistical analysis of observational data,
only the amplitude $A_1$ is constrained,
and there is no constraint on the index $\alpha$.

Making use of these resulting amplitude constraints  from experiment
as well as the calculated strain spectrum $h_c(f)$,
we will be able to
put constraint on the index $\alpha$,
and thus on the inflation index $\beta=\alpha-1$.
To do this, in Fig.\ref{a-alpha} we plot the calculated $h_c(f)$
in Eq.(\ref{strain}) at fixed frequency $f=1/(1\,  \rm{yr})$,
i.e, $h_c(\rm{yr}^{-1})$, as a function of $\alpha$,
where three cases of the ratio $r=0.01$, $0.1$ and $0.26$ are given, respectively.
Note that the upper limit curves $A_1(\alpha)$
intersects that of the calculated $h_c(\rm{yr}^{-1})$.
These intersection points give rise to
the corresponding constraints on the index $\alpha$.
The intersection point can be obtained by
setting equal the right hand sides of Eq.(\ref{strain}) and Eq.(\ref{hc1}),
yielding following constraint equation£º
\be\label{pulsar}
  \frac{\Delta_\mathcal{R}(k_0)r^{1/2}}{\sqrt{2}}
  \left(\frac{{\rm yr^{-1}}}{f_0}\right)^{\alpha}
\left(\frac{f_0}{f_2}\right)=A_1(\alpha).
\ee
The solution of this equation gives the upper limit on $\alpha$.
In Table 1, the resulting constraints of $\alpha$,
and $\beta$ as well,
given by  the above equation  are listed,
where we have presented the results imposed by various PTA groups,
and for values of the ratio  $r=0.26,\,  0.1,\,  0.01$, respectively.
It is worth to point out that the results given in Table 1 are little different from
those shown in  \cite{Jenet}. The difference is induced by many reasons.
Firstly, the normalized spectral amplitude in our results depends on the tensor-to-scalar ratio $r$. Secondly, the pivot frequency in this paper is different from that in Ref.\cite{Jenet}.
 We chose the pivot frequency to be $f_0$, while the pivot frequency was chosen to be $H_0$ in Ref.\cite{Jenet}. Thirdly, here we considered  the acceleration stage of the universe.  On the other hand, the estimation of $a_2/a_H\simeq10^{-4}$ in \cite{Jenet} is a little bit large. We find that the constraints of $\alpha$ or $\beta$ depend on this value sensitively.
These resulting upper limits of $\alpha$ and $\beta$ are quite large,
telling that the constraints by PTAs are not as stringent as those
by BBN, CMB and  LIGO \cite{zhangtong}. However, the constraints of $\alpha$ and $\beta$ are tighter than those shown in \cite{zhaozhang}. Hence, the PTAs give  more strict constraints on the non-standard RGW model.

\begin{figure}
\resizebox{100mm}{!}{\includegraphics{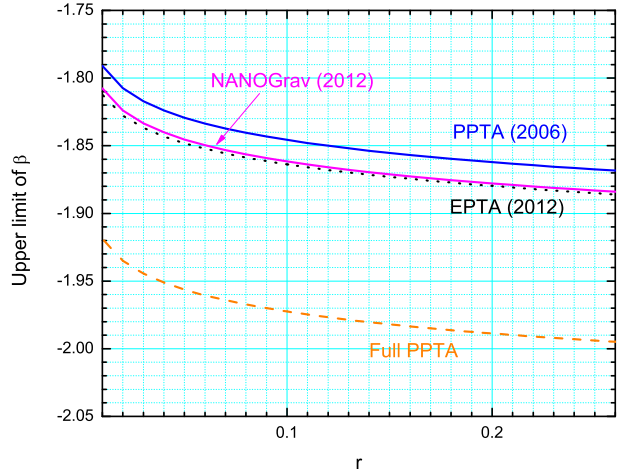}}
\caption{\label{beta-r}
The upper limit  of $\beta$ varies with $r$ according to Eq.(\ref{pulsar})
for various PTAs without the quantum normalization.
}
\end{figure}

Notice that,  even though the  parameters $\beta$ and $r$ initially independent
in a generic theoretical model,
the constraint equation (\ref{pulsar}) now imposes a relation between $r$ and the upper
limit of $\alpha$.
This relation is not from the quantum normalization mentioned earlier,
but, instead, from the constraints of PTAs in combination with a specific RGWs model.
We plot the resulting upper limit of
$\beta= \alpha-1$ as a function of $r$
for various PTA groups in Fig. \ref{beta-r}.
It shows that,
a smaller upper limit  $\beta$  is associated with a larger value of $r$.

\begingroup
\begin{table}
\caption{\label{table}
The upper limits of $\alpha$ with different values of $r$ given by different PTA groups.
}
\begin{center}
\begin{tabular}{|l|c|c|c|}
\hline
 \quad PTA groups \quad &\quad $r=0.26$ \quad&\quad $r=0.1$ \quad&\quad $r=0.01$   \\
\hline\hline
\quad PPTA(2006) \cite{Jenet} \quad  &\,$\alpha\leq-0.87\ (\beta\leq-1.87)$ &  \,$\alpha\leq-0.85\ (\beta\leq-1.85)$ &  \,$\alpha\leq-0.79\ (\beta\leq-1.79)$    \\
\hline
\quad EPTA(2012) \cite{Haasteren} \quad  &\,$\alpha\leq-0.89\ (\beta\leq-1.89)$ &\,$\alpha\leq-0.86\ (\beta\leq-1.86)$&\,$\alpha\leq-0.81\ (\beta\leq-1.81)$  \\
\hline
\quad NANOGrav(2012) \cite{Demorest} \quad   &\,$\alpha\leq-0.88\ (\beta\leq-1.88)$ &\,$\alpha\leq-0.86\ (\beta\leq-1.86)$&\,$\alpha\leq-0.81\ (\beta\leq-1.81)$ \\
\hline
\quad Full PPTA \cite{Jenet}\quad  &\,$\alpha\leq-0.99\ (\beta\leq-1.99)$&\,$\alpha\leq-0.97\ (\beta\leq-1.97)$&\,$\alpha\leq-0.92\ (\beta\leq-1.92)$  \\
\hline
\end{tabular}
\end{center}
\end{table}
\endgroup

\section{ Constraints by the quantum normalization }

\begin{figure}
\resizebox{80mm}{!}{\includegraphics{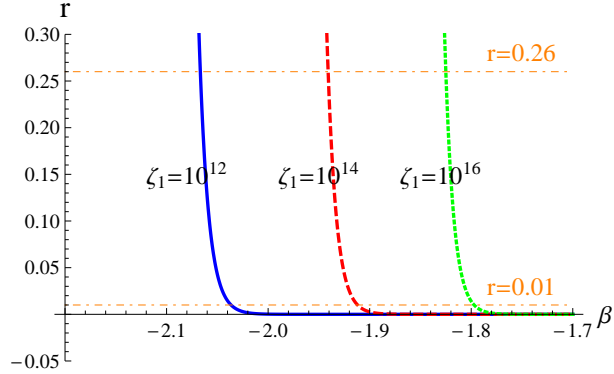}}
\caption{\label{rbeta}
The quantum normalization predicts
a relation between $r$ and $\beta$
for various values of $\zeta_1$.
The upper dashed line represents the constraint of
$r$ given by Planck \cite{Planck2} and
the lower dashed line represents solid line represents
lower limit of $r\simeq 0.01$
required by the model-independent measure of the degree
of fine-tuning \cite{Boyle}.
}
\end{figure}

In general, from  theoretical perspective,
 $\beta$ and $r$ could be independent parameters,
when RGWs were generated by a generic mechanism in the early universe.
However,  there is another possible way in which
 $\beta$ and $r$ are related.
If one treats the RGWs field  $h_{ij}$ as a quantum field,
and requires that, when initially generated,
it be in the vacuum state
with an energy $\frac{1}{2}\hbar \omega$ for each mode $k=\omega/2\pi$,
one ends up with the so-called quantum normalization for the amplitude
of RGWs \cite{grishchuk}:
\be
h(k,\tau_i)=8\sqrt{\pi}/\lambda_i.
\ee
During the inflationary expansion,
 the reduced wavelength of
each wave mode crossed the horizon when $\lambda_i/(2\pi)=1/H(\tau_i)$, which
leads to
\be\label{aini}
 A=  \frac{4bl_{Pl}}{\sqrt{\pi}l_0},
\ee
where $l_{Pl}$ is the Plank length,
 $b\equiv\gamma^{2+\beta}/|1+\beta|^{1+\beta}$,
 and
\be\label{l0}
  l_0=b H_0^{-1} \zeta_1^{\frac{\beta-\beta_s}{1+\beta_s}}
  \zeta_s^\beta \zeta_2^{\frac{\beta-1}{2}}
  \zeta_E^{-(1+\frac{1+\beta}{\gamma})},
\ee
 in our notation \cite{Miao}.
Let us explore the consequences of  this normalization.
 From Eqs.(\ref{primordial}), (\ref{aini}) and (\ref{l0}),
 one obtains the following relation:
\be\label{arelation}
\Delta_\mathcal{R}(k_0)r^{1/2}\left(\frac{k_H}{k_0}\right)^\beta
\zeta_E^{\frac{\gamma+2}{\gamma}}=4\pi^{-\frac{1}{2}}l_{Pl}H_0
\zeta_1^{\frac{\beta_s-\beta}{1+\beta_s}}\zeta_s^{-\beta}
\zeta_2^{\frac{1-\beta}{2}}
  \zeta_E^{(1+\frac{1+\beta}{\gamma})},
\ee
which tells that, according to the quantum normalization,
 the two major parameters $r$ and $\beta$ are
no longer independent,
but rather related.
In particular,
$r$ and $\beta$ are related  by a function $r=r(\beta)$
in a specific form,
when other parameters such as
$\beta_s$, $\gamma$, $\zeta_1$, $\zeta_2$, and $\zeta_s$ are all held fixed.
In Fig. \ref{rbeta} we plot $r=r(\beta)$ for fixed $\trh=10^7$ GeV,
showing that
 a smaller $\beta$ is associated with a greater $r$.
This pattern of behavior predicted the quantum normalization
is formally similar to that in Fig. \ref{beta-r} from PTAs.
Besides,
Fig. \ref{rbeta} also demonstrates  that
when the parameter $\zeta_1$ of the reheating
is allowed to vary,
a larger $\zeta_1$ shifts the curve $r=r(\beta)$  to
a larger $\beta$.

In the scenario of generation of RGWs
as perturbations of metric during inflation,
there is a theoretical upper limit on the index $\beta$ \cite{grishchuk,zhang2}.
During the inflationary expansion,
the wavelength of
each mode of the gravitational waves the horizon-crossing, $\lambda_i/(2\pi)=1/H(\tau_i)$,
should be greater than the Plank length, i.e., $\lambda_i>l_{Pl}$,
which is a reasonable requirement for validity of treatment
of the spacetime during inflation as classical.
With the help of Eqs. (\ref{inflation}),
the wavelength at the crossing
is written as $\lambda_i=2\pi\frac{l_0}{b}(\frac{f_H}{f})^{2+\beta}$.
So, the requirement becomes
\be\label{betalimit}
  \left(\frac{f}{f_H}\right)^{2+\beta}<\frac{8\sqrt{\pi}}{A},
\ee
where $A$ is the amplitude from the quantum normalization in Eq.(\ref{aini}).
The rate of the primordial nucleosynthesis requires
an upper bound of frequency  \cite{grishchuk,Miao},
which is $f_1\simeq4\times10^{10}$ Hz
when the effect of dark energy is included  \cite{Tong6}.
Substituting this bound into Eq. (\ref{betalimit})
and using Eq.(\ref{primordial}),
one obtains the theoretical upper limit on the index
$\beta<-1.87$ for $r=0.26$,
and $\beta<-1.85$ for $r=0.01$, respectively.

It is interesting to notice that
the  upper limits of $\beta$ imposed by the quantum normalization
are comparable to those by the current PTAs in Table 1
in the last section.
So the result of the current PTAs experiments
supports the scenario that
the wavelengths $\lambda$ of RGWs
were much greater than the Planck length $l_{Pl}$
when the wave modes crossed over the horizon during inflation.

\section{ Constraints upon reheating by both PTAs and quantum normalization}

In the above analysis on the parameters $\beta$ and $r$,
we have fixed other remaining parameters of RGWs, such as $\zeta_1$, and etc.
In fact, using the PTAs results, in combination with the quantum normalization,
one can also put constraints on certain other parameters of RGWs.
Now we focus on the parameter $\zeta_1$ for the reheating,
a very important parameter for the cosmology.
Although  its value can be predicted by certain inflation models \cite{Mielczarek},
but so far it is not constrained by experiment other than WMAP on CMB.

\begin{figure}
\resizebox{100mm}{!}{\includegraphics{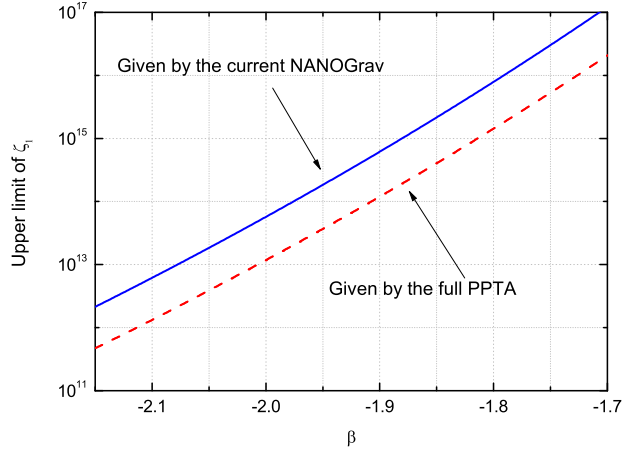}}
\caption{\label{zeta1-r}
The upper limit of $\zeta_1$ given
by different PTAs  under the quantum normalization.
}
\end{figure}

Again,
using Eq.(\ref{arelation}) now with both $\beta$ and $\zeta_1$ as being free,
one has $r$ as a function: $r=r(\beta,\zeta_1)$.
Substituting this into Eq. (\ref{strain}) at  $f=1/(1\,\rm{yr})$
yields the strain amplitude $h_{c}$ as a function of both $\beta$ and $\zeta_1$:
$h_{c} =  h_{c}(\beta,\zeta_1)$.
Requiring that this be not higher than the amplitude $A_1(\beta)$ from PTAs,
\be
h_{c}(\beta,\zeta_1)\leq A_1(\beta),
\ee
one can get an upper limit of $\zeta_1$ for each $\beta$.
The results are plotted in Fig.\ref{zeta1-r}
with  the upper limits of $\zeta_1$ as a function of $\beta$,
where we have demonstrated for two sets of constraints of PTAs.
The future full PPTA
will give stronger constraints than the current PTAs.
For the  value $\beta=-2.02$,
the upper bound is $\zeta_1< 3.6\times 10^{13}$ by the current NANOGrav,
and
the full PPTA will give a bound $\zeta_1< 7.6\times 10^{12}$. For the
value $\beta=-2$ corresponding to the de Sitter expansion of inflation,
the current NANOGrav and the full PPTA give the bounds  $\zeta_1< 5.7\times 10^{13}$
and $\zeta_1< 1.2\times 10^{13}$, respectively.
Therefore, to the very early cosmic processes,
the future full PTPA can also serve as an important probe,
complementary to others, such as CMB, etc.

\begin{figure}
\resizebox{100mm}{!}{
\includegraphics{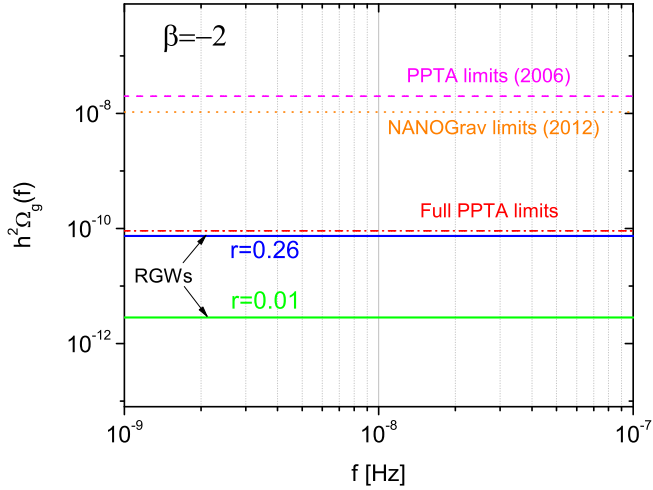}}
\caption{\label{relic-2}
The upper limits on the energy density spectrum $\Omega_g(f)h^2$
given by various PTAs compared with the theoretical  $\Omega_g(f)h^2$
 with $r=0.26$ and $r=0.01$, respectively.
Here $\beta=-2$ is taken so that $\Omega_g(f)h^2$ is flat.
}
\end{figure}

\section{ Conclusions and Discussions }

Our analysis  demonstrated constraints on RGWs given by different   PTAs,
 which can be expressed
in terms of the upper bounds of the spectral index $\beta$ determined by inflation.
We find that  the PTAs give  more strict constraints on the non-standard RGW model than the standard RGW model.
On the other hand, the requirement of the quantum normalization also
imposes constraints on RGWs.
These two sets of bounds are comparable, and consistent  to each other.
But  the current PTAs bounds are not as stringent as those by LIGO and VIRGO.
However,  as an advantage,
when the combination of the PTAs and the quantum normalization
are both applied to RGWs,
constraints  can be obtained on the parameter $\zeta_1$ of the reheating,
an important process of the very early universe,
which is currently less understood than the inflation.

\begin{figure}
\resizebox{100mm}{!}{\includegraphics{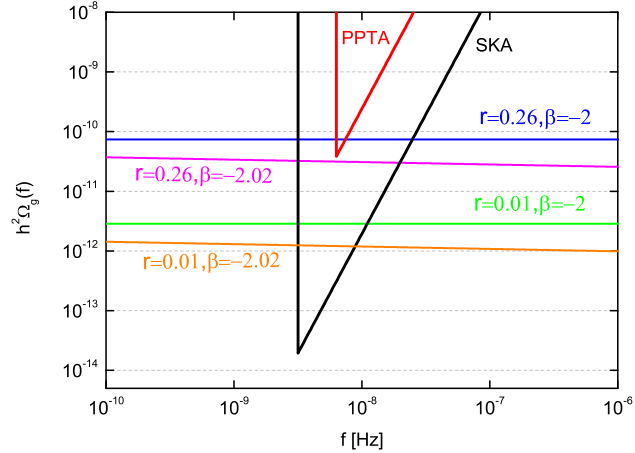}}
\caption{\label{8}
The energy density spectrum $\Omega_g(f)h^2$
to be explored by full PPTA \cite{Jenet}
and the planned SKA.
Here the probing of SKA is converted from
that from Ref.\cite{Sesana},
assuming the monitoring
of $20$ pulsars for $10$ yr at a precision level of $\delta t_{rms}\sim 10$ ns.
}
\end{figure}

To examine the possibility of detection of RGWs by PTAs, We just take $\beta=-2$ as
an example. In Fig.\ref{relic-2} we plot
the upper limit of the energy density spectrum $\Omega_g(f) h^2$
set by
the PPTA  \cite{Jenet},
the  current  NANOGrav \cite{Demorest},
and the full PPTA \cite{Jenet}, respectively.
For comparison,
we also plot the theoretical $\Omega_gh^2$ for $r=0.26$ and for $r=0.01$,
respectively.
The curves for $\Omega_g(f)h^2$ in Fig.\ref{relic-2} are
flat,
since, for the case $\beta=-2$, by Eq.(\ref{omega}) and Eq.(\ref{strain}),
the function $\Omega_g(f)h^2$ is independent of frequency $f$.
It is clearly seen that
the current PTAs  fall short of  at least two orders of magnitude,
but the full PPTA will nearly catch the signal of  the  RGWs of $r=0.26$.
On the other hand, if $r$ be a much smaller value, even the full PPTA will have no chance
 to detect  RGWs.
However,
the future instruments such as  SKA \cite{Kramer}
will have enormously improved sensitivity,
and would have a greater potential to detect  RGWs.
Fig.\ref{8} shows the possible probing of SKA
and the theoretical $\Omega_g(f)h^2$ for $\beta=-2$ and $\beta=-2.02$,
respectively.
One sees that the sensitivity of  SKA would be enough to detect  RGWs.
Furthermore, given such a capability,
SKA will also be able to put constrains upon the parameter $\zeta_1$
of the reheating  stringent than PTAs. Therefore, SKA will become a powerful
tool to detect or constrain RGWs and other kinds of stochastic gravitational wave background.

All the analysis and results presented in this paper
are arrived under the assumption that
the tensorial running index $\alpha_t= 0$.
The curvature of the inflation potential determines $\alpha_t$.
Inflation models with large values of $\alpha_t$
predict a spectrum of RGWs  tilting up on short wavelengths,
increasing the chance to detection, and vice versa.


Aside RGWs,
there is a stochastic background of gravitational waves
generated by supermassive
  black hole binaries \cite{Jaffe,Sesana},
whose power-law spectral index $\alpha=-2/3$,
different from  $\alpha\simeq-1$ of RGWs.
With a relatively wider range of detection frequencies,
SKA might be able to distinguish the two different kinds of
gravitational wave background.

~
~
\

{ACKNOWLEDGMENT}: M.L. Tong's work has been supported by the NSFC  under Grant No. 11103024
and  the  program of the light in China's Western Region of CAS.
Y. Zhang's work has been supported by
the NSFC under Grant No.11073018, 11275187,   SRFDP, and CAS.
W. Zhao's work has been supported by the NSFC under Grant No. 11173021, 11075141 and project
of Knowledge Innovation Program of Chinese Academy of Science.
J.Z. Liu's work has been supported by NSFC  under Grant No. 11103054,
the  program of the light in China¡¯s Western Region of CAS, XNFC,
and  ¡°Beyond the Horizons¡± of
National Astronomical Observatories and John Templeton Foundation.

\end{document}